\documentclass[12pt]{article}
\usepackage{amssymb,amscd,amsmath, array}
\catcode `\@=11
\@addtoreset{equation}{section}

\newtheorem{thm}{Theorem}[section]

\newcommand{\be}{\begin{equation}}
\newcommand{\ee}{\end{equation}}
\newcommand{\bea}{\begin{eqnarray}}
\newcommand{\eea}{\end{eqnarray}}
\begin{document}
\begin{titlepage}

\begin{center}
{\bf \Large{Yang-Mills Models in the Causal Approach: Perturbation Theory up to the Second Order\\}}
\end{center}
\vskip 1.0truecm
\centerline{D. R. Grigore, 
\footnote{e-mail: grigore@theory.nipne.ro}}
\vskip5mm
\centerline{Department of Theoretical Physics, Institute for Physics and Nuclear
Engineering ``Horia Hulubei"}
\centerline{Institute of Atomic Physics}
\centerline{Bucharest-M\u agurele, P. O. Box MG 6, ROM\^ANIA}

\vskip 2cm
\bigskip \nopagebreak
\begin{abstract}
\noindent
We consider the standard model up to the second order of the perturbation theory (in the causal approach) 
and derive the most general form of the interaction Lagrangian for an arbitrary number of Higgs fields.
\end{abstract}
%\newpage\setcounter{page}1
\end{titlepage}

\section{Introduction}

Renormalization theory consists in the construction of scattering matrix in the perturbative sense; this amounts to
the construction of the chronological products such that Bogoliubov axioms are verified \cite{BS},
\cite{EG}, \cite{DF}; for every set of Wick monomials 
$ 
W_{1}(x_{1}),\dots,W_{n}(x_{n}) 
$
acting in the Fock space
$
{\cal H}
$
one associates the distribution-valued operators
$ 
T^{W_{1},\dots,W_{n}}(x_{1},\dots,x_{n}) 
\equiv
T(W_{1}(x_{1}),\dots,W_{n}(x_{n})) 
$ 
called chronological products. The existence of the chronological products had been established by Epstein-Glaser \cite{EG}, \cite{Gl}
in a purely constructive way: if one knows them up to the order $n - 1$ then one can construct them in order $n$
using distribution splitting techniques; alternatively one can use the procedure of extension of distributions \cite{Sto1}. 
The procedure does not fix uniquely these products but there are some natural 
limitation on the arbitrariness. If the arbitrariness does not grow with $n$ we
have a renormalizable theory. 

Gauge theories can be understood in this framework if we known how to describe particles of higher spin. If the quantization procedure
is not cleverly chosen then such theories are not renormalizable. One can save renormalizablility using ghost fields.
Such theories are defined in a Fock space
$
{\cal H}
$
with indefinite metric, generated by physical and un-physical fields (called
{\it ghost fields}). One selects the physical states assuming the existence of
an operator $Q$ called {\it gauge charge} which verifies
$
Q^{2} = 0
$
and such that the {\it physical Hilbert space} is by definition
$
{\cal H}_{\rm phys} \equiv Ker(Q)/Im(Q).
$
One can define a grading of the Hilbert(called ghost number) space such that $Q$ has ghost number $1$ and it
raises the ghost number of any state.
 
A gauge theory assumes also that there exists a Wick polynomial of null ghost
number
$
T(x)
$
called {\it the interaction Lagrangian} such that
\be
~[Q, T] = i \partial_{\mu}T^{\mu}
\label{gau1}
\ee
for some other Wick polynomials
$
T^{\mu}.
$
This relation means that the expression $T$ leaves invariant the physical
states, at least in the adiabatic limit. Indeed, if this is true we have:
\be
T(f)~{\cal H}_{\rm phys}~\subset~~{\cal H}_{\rm phys}  
\label{gau2}
\ee
up to terms which can be made as small as desired (making the test function $f$
flatter and flatter). In all known models one finds out that there exist a chain
of Wick polynomials
$
T^{\mu},~T^{\mu\nu},~T^{\mu\nu\rho},\dots
$
such that:
\be
~[Q, T] = i \partial_{\mu}T^{\mu}, \quad
[Q, T^{\mu}] = i \partial_{\nu}T^{\mu\nu}, \quad
[Q, T^{\mu\nu}] = i \partial_{\rho}T^{\mu\nu\rho},\dots
\label{descent}
\ee
It so happens that for all these models the expressions
$
T^{\mu\nu},~T^{\mu\nu\rho},\dots
$
are completely antisymmetric in all indexes; it follows that the chain of
relation stops at the step $4$ (if we work in four dimensions). We can also use
a compact notation
$
T^{I}
$
where $I$ is a collection of indexes
$
I = [\nu_{1},\dots,\nu_{p}]~(p = 0,1,\dots,)
$
and the brackets emphasize the complete antisymmetry in these indexes. All these
polynomials have the same canonical dimension
\be
\omega(T^{I}) = \omega_{0},~\forall I
\ee
and because the ghost number of
$
T \equiv T^{\emptyset}
$
is supposed null, then we also have:
\be
gh(T^{I}) = |I|.
\ee
One can write compactly the relations (\ref{descent}) as follows:
\be
d_{Q}T^{I} - i~\partial_{\mu}T^{I\mu} = 0.
\label{descent1}
\ee

Now we can construct the chronological products
$$
T^{I_{1},\dots,I_{n}}(x_{1},\dots,x_{n}) \equiv
T(T^{I_{1}}(x_{1}),\dots,T^{I_{n}}(x_{n}))
$$
according to the general procedure and generalize in a natural way the preceding relation. We consider arbitrary co-chains
of the form
$$
C^{I_{1},\dots,I_{n}}(x_{1},\dots,x_{n})
$$
with appropriate symmetry properties \cite{cohomology} and define the operator
\be
\delta C^{I_{1},\dots,I_{n}} \equiv
\sum_{l=1}^{n} (-1)^{s_{l}} \frac{\partial}{\partial x^{\mu}_{l}}
C^{I_{1},\dots,I_{l}\mu,\dots,I_{n}}
\label{delta}
\ee
where
\be
s_{l} \equiv \sum_{j=1}^{l-1} |I|_{j}
\ee
and we define the {\it BRST operator}
\be
s \equiv d_{Q} - i~\delta;
\ee
similarly we define the {\it anti-BRST operator}:
\be
\bar{s} \equiv d_{Q} - i~\delta
\ee
and note that
\be
s\bar{s} = \bar{s}s = 0. 
\ee

We say that the theory is gauge invariant in all orders of the perturbation theory if the following set of identities
generalizing (\ref{descent1}):
\be
sT^{I_{1},\dots,I_{n}} \equiv d_{Q}T^{I_{1},\dots,I_{n}} -
i \sum_{l=1}^{n} (-1)^{s_{l}} \frac{\partial}{\partial x^{\mu}_{l}}
T^{I_{1},\dots,I_{l}\mu,\dots,I_{n}} = 0.
\label{gauge}
\ee

In particular, the case
$
I_{1} = \dots = I_{n} = \emptyset
$
it is sufficient for the gauge invariance of the scattering matrix, at least
in the adiabatic limit: we have the same argument as for relation (\ref{gau2}).
Such identities can be usually broken by {\it anomalies} i.e. expressions of the
type
$
A^{I_{1},\dots,I_{n}}
$
which are quasi-local and might appear in the right-hand side of the relation
(\ref{gauge}). If one eliminates the anomalies, some restrictions must be
imposed on the interaction Lagrangian, besides those following from
(\ref{gau1}).

We call {\it co-cycles} the co-chains verifying
$
sC = 0.
$
A special kind of co-cycles are the {\it co-boundaries} namely expressions of the type
$
C = \bar{s}B
$
for an arbitrary co-chain $B$.

In this paper we consider all these restrictions up to the second order of the
perturbation theory and determine the most general form for $T$. This problem
was previously analyzed in great detail in \cite{Sc2}, but no general solution
was found. For the standard model of the electro-weak interactions 
we find three type of solutions corresponding to the ratio
$
\gamma \equiv \frac{m_{Z}~\cos\theta}{m_{W}}
$
taking the values $0$, $> 0$ and $< 0$ respectively. The first case is 
relevant for the usual standard model.

In the next Section we remind the construction of the Fock space for a general
Yang-Mills theory, mainly to fix the notations. In Section \ref{pert} we give the 
some necessary facts about perturbation theory in the causal approach. The study 
of first and second order of the perturbation theory is done in Section \ref{Lym}.

\newpage
\section{The Cohomology of the Gauge Charge Operator\label{q}}

To fix the notations, we give in the first five subsections a short account of the description 
of Yang-Mills fields in the causal formalism following \cite{cohomology}.

\subsection{Massless Particles of Spin $1$ (Photons)}

We consider a vector space 
$
{\cal H}
$
of Fock type generated (in the sense of Borchers theorem) by the vector field 
$
v_{\mu}
$ 
(with Bose statistics) and the scalar fields 
$
u, \tilde{u}
$
(with Fermi statistics). The Fermi fields are usually called {\it ghost fields}.
We suppose that all these (quantum) fields are of null mass. Let $\Omega$ be the
vacuum state in
$
{\cal H}.
$
In this vector space we can define a sesquilinear form 
$<\cdot,\cdot>$
in the following way: the (non-zero) $2$-point functions are by definition:
\bea
<\Omega, v_{\mu}(x_{1}) v_{\mu}(x_{2})\Omega> =i~\eta_{\mu\nu}~D_{0}^{(+)}(x_{1}
- x_{2}),
\nonumber \\
<\Omega, u(x_{1}) \tilde{u}(x_{2})\Omega> =- i~D_{0}^{(+)}(x_{1} - x_{2})
\qquad
<\Omega, \tilde{u}(x_{1}) u(x_{2})\Omega> = i~D_{0}^{(+)}(x_{1} - x_{2})
\eea
and the $n$-point functions are generated according to Wick theorem. Here
$
\eta_{\mu\nu}
$
is the Minkowski metrics (with diagonal $1, -1, -1, -1$) and 
$
D_{0}^{(+)}
$
is the positive frequency part of the Pauli-Jordan distribution
$
D_{0}
$
of null mass. To extend the sesquilinear form to
$
{\cal H}
$
we define the conjugation by
\be
v_{\mu}^{\dagger} = v_{\mu}, \qquad 
u^{\dagger} = u, \qquad
\tilde{u}^{\dagger} = - \tilde{u}.
\ee

Now we can define in 
$
{\cal H}
$
the operator $Q$ according to the following formulas:
\bea
~[Q, v_{\mu}] = i~\partial_{\mu}u,\qquad
[Q, u] = 0,\qquad
[Q, \tilde{u}] = - i~\partial_{\mu}v^{\mu}
\nonumber \\
Q\Omega = 0
\label{Q-0}
\eea
where by 
$
[\cdot,\cdot]
$
we mean the graded commutator. One can prove that $Q$ is well defined. Indeed,
we have the causal commutation relations 
\be
~[v_{\mu}(x_{1}), v_{\mu}(x_{2}) ] =i~\eta_{\mu\nu}~D_{0}(x_{1} - x_{2})~\cdot
I,
\qquad
[u(x_{1}), \tilde{u}(x_{2})] = - i~D_{0}(x_{1} - x_{2})~\cdot I
\ee
and the other commutators are null. The operator $Q$ should leave invariant
these relations, in particular 
\be
[Q, [ v_{\mu}(x_{1}),\tilde{u}(x_{2})]] + {\rm cyclic~permutations} = 0
\ee
which is true according to (\ref{Q-0}). It is useful to introduce a grading in 
$
{\cal H}
$
as follows: every state which is generated by an even (odd) number of ghost
fields and an arbitrary number of vector fields is even (resp. odd). We denote
by 
$
|f|
$
the ghost number of the state $f$. We notice that the operator $Q$ raises the
ghost number of a state (of fixed ghost number) by an unit. The usefulness of
this construction follows from:
\begin{thm}
The operator $Q$ verifies
$
Q^{2} = 0.
$ 
The factor space
$
Ker(Q)/Ran(Q)
$
is isomorphic to the Fock space of particles of zero mass and helicity $1$
(photons). 
\end{thm}

Let us check that the gauge structure above gives the right physical Hilbert space at least in the one-particle Hilbert space. 
The generic form of a state 
$
\Psi \in {\cal H}^{(1)} \subset {\cal H}
$
from the one-particle Hilbert subspace is
\be
\Psi = \left[ \int f_{\mu}(x) v^{\mu}(x) + \int g_{1}(x) u(x) + \int g_{2}(x) \tilde{u}(x) \right] \Omega
\ee
with test functions
$
f_{\mu}, g_{1}, g_{2}
$
verifying the wave equation equation. We impose the condition 
$
\Psi \in Ker(Q) \quad \Longleftrightarrow \quad Q\Psi = 0;
$
we obtain 
$
\partial^{\mu}f_{\mu} = 0
$
and
$
g_{2} = 0
$
i.e. the generic element
$
\Psi \in {\cal H}^{(1)} \cap Ker(Q)
$
is
\be
\Psi = \left[ \int f_{\mu}(x) v^{\mu}(x) + \int g(x) u(x) \right] \Omega
\label{kerQ}
\ee
with $g$ arbitrary and 
$
f_{\mu}
$
constrained by the transversality condition 
$
\partial^{\mu}f_{\mu} = 0;
$
so the elements of
$
{\cal H}^{(1)} \cap Ker(Q)
$
are in one-one correspondence with couples of test functions
$
(f_{\mu}, g)
$
with the transversality condition on the first entry. Now, a generic element
$
\Psi^{\prime} \in {\cal H}^{(1)} \cap Im(Q)
$
has the form 
\be
\Psi^{\prime} = Q\Phi = \left[ - \int \partial^{\mu}f^{\prime}_{\mu}(x) u(x) 
+ \int \partial_{\mu}g^{\prime}(x) v^{\mu}(x) \right] \Omega
\label{imQ}
\ee
so if
$
\Psi \in {\cal H}^{(1)} \cap Ker(Q)
$
is indexed by
$
(f_{\mu}, g)
$
then 
$
\Psi + \Psi^{\prime}
$
is indexed by 
$
(f_{\mu} + \partial_{\mu}g^{\prime}, g - \partial^{\mu}f^{\prime}_{\mu}).
$
If we take 
$
f^{\prime}_{\mu}
$
conveniently we can make 
$
g = 0.
$
We introduce the equivalence relation 
$
f_{\mu}^{(1)} \sim f_{\mu}^{(2)} \quad \Longleftrightarrow 
f_{\mu}^{(1)} - f_{\mu}^{(2)} = \partial_{\mu}g^{\prime}
$
and it follows that the equivalence classes from
$
Ker(Q)/Im(Q)
$ 
are indexed by equivalence classes of wave functions
$
[f_{\mu}];
$
we have obtained the usual one-particle Hilbert space for the photon. One can generalize this argument for the multi-particle
Hilbert space \cite{cohomology}. 
%\newpage

\subsection{Massive Particles of Spin $1$ (Heavy Bosons)}

We repeat the whole argument for the case of massive photons i.e. particles of
spin $1$ and positive mass. 

We consider a vector space 
$
{\cal H}
$
of Fock type generated (in the sense of Borchers theorem) by the vector field 
$
v_{\mu},
$ 
the scalar field 
$
\Phi
$
(with Bose statistics) and the scalar fields 
$
u, \tilde{u}
$
(with Fermi statistics). We suppose that all these (quantum) fields are of mass
$
m > 0.
$
In this vector space we can define a sesquilinear form 
$<\cdot,\cdot>$
in the following way: the (non-zero) $2$-point functions are by definition:
\bea
<\Omega, v_{\mu}(x_{1}) v_{\mu}(x_{2})\Omega> =i~\eta_{\mu\nu}~D_{m}^{(+)}(x_{1}
- x_{2}),
\quad
<\Omega, \Phi(x_{1}) \Phi(x_{2})\Omega> =- i~D_{m}^{(+)}(x_{1} - x_{2})
\nonumber \\
<\Omega, u(x_{1}) \tilde{u}(x_{2})\Omega> =- i~D_{m}^{(+)}(x_{1} - x_{2}),
\qquad
<\Omega, \tilde{u}(x_{1}) u(x_{2})\Omega> = i~D_{m}^{(+)}(x_{1} - x_{2})
\eea
and the $n$-point functions are generated according to Wick theorem. Here
$
D_{m}^{(+)}
$
is the positive frequency part of the Pauli-Jordan distribution
$
D_{m}
$
of mass $m$. To extend the sesquilinear form to
$
{\cal H}
$
we define the conjugation by
\be
v_{\mu}^{\dagger} = v_{\mu}, \qquad 
u^{\dagger} = u, \qquad
\tilde{u}^{\dagger} = - \tilde{u},
\qquad \Phi^{\dagger} = \Phi.
\ee

Now we can define in 
$
{\cal H}
$
the operator $Q$ according to the following formulas:
\bea
~[Q, v_{\mu}] = i~\partial_{\mu}u,\qquad
[Q, u] = 0,\qquad
[Q, \tilde{u}] = - i~(\partial_{\mu}v^{\mu} + m~\Phi)
\qquad
[Q,\Phi] = i~m~u,
\nonumber \\
Q\Omega = 0.
\label{Q-m}
\eea
One can prove that $Q$ is well defined. We have a result similar to the first
theorem of this Section:
\begin{thm}
The operator $Q$ verifies
$
Q^{2} = 0.
$ 
The factor space
$
Ker(Q)/Ran(Q)
$
is isomorphic to the Fock space of particles of mass $m$ and spin $1$ (massive
photons). 
\end{thm}

The proof is similar to the massless case.

%\newpage
\subsection{The Generic Yang-Mills Case}

The situations described above (of massless and massive photons) are susceptible
of the following generalizations. We can consider a system of 
$
r_{1}
$ 
species of particles of null mass and helicity $1$ if we use in the first part
of this Section 
$
r_{1}
$ 
triplets
$
(v^{\mu}_{a}, u_{a}, \tilde{u}_{a}), a \in I_{1}
$
of massless fields; here
$
I_{1} = I_{\rm null~mass}
$
is a set of indexes of cardinal 
$
r_{1}.
$
All the relations have to be modified by appending an index $a$ to all these
fields. 

In the massive case we have to consider 
$
r_{2}
$ 
quadruples
$
(v^{\mu}_{a}, u_{a}, \tilde{u}_{a}, \Phi_{a}),  a \in I_{2}
$
of fields of mass 
$
m_{a}
$; here
$
I_{2} = I_{\rm positive~mass}
$
is a set of indexes of cardinal 
$
r_{2}.
$

We can consider now the most general case involving fields of spin not greater
that $1$.
We take 
$
I = I_{1} \cup I_{2} \cup I_{3}
$
a set of indexes and for any index we take a quadruple
$
(v^{\mu}_{a}, u_{a}, \tilde{u}_{a},\Phi_{a}), a \in I
$
of fields with the following conventions:
(a) For
$
a \in I_{1}
$
we impose 
$
\Phi_{a} = 0
$
and we take the masses to be null
$
m_{a} = 0;
$
(b) For
$
a \in I_{2}
$
we take the all the masses strictly positive:
$
m_{a} > 0;
$
(c) For 
$
a \in I_{3} = I_{\rm Higgs}
$
we take 
$
v_{a}^{\mu}, u_{a}, \tilde{u}_{a}
$
to be null and the fields
$
\Phi_{a} \equiv \phi^{H}_{a} 
$
of mass 
$
m_{a} \equiv m^{H}_{a} \geq 0.
$
The fields
$
\phi^{H}_{a} 
$
are called {\it Higgs fields}.

If we define
$
m_{a} = 0, \forall a \in I_{3}
$
then we can define in 
$
{\cal H}
$
the operator $Q$ according to the following formulas for all indexes
$
a \in I:
$
\bea
~[Q, v^{\mu}_{a}] = i~\partial^{\mu}u_{a},\qquad
[Q, u_{a}] = 0,
\nonumber \\
~[Q, \tilde{u}_{a}] = - i~(\partial_{\mu}v^{\mu}_{a} + m_{a}~\Phi_{a})
\qquad
[Q,\Phi_{a}] = i~m_{a}~u_{a},
\nonumber \\
Q\Omega = 0.
\label{Q-general}
\eea

If we consider matter fields also i.e some set of Dirac fields with Fermi
statistics
$
\Psi_{A}
$ 
of mass
$
M_{A}, A \in I_{4} = I_{\rm Dirac}
$
and we impose
\be
d_{Q}\Psi_{A} = 0 
\ee
and the space 
$
{\cal P}_{0}
$
is generated by
$
\Psi_{A}
$
and
$
\bar{\Psi}_{A}
$
also. We denote by $M$ the (diagonal) mass matrix of the Dirac Fields
\be
M_{AB} = \delta_{AB}~M_{A}.
\ee
\newpage

\section{Perturbation Theory\label{pert}}
We provide the necessary elements of (second order) of perturbation theory. Formally, 
we want to compute the scattering matrix
\be
S(g) \equiv I + i \int dx g(x) T(x) 
\nonumber \\
+ \frac{i^{2}}{ 2} \int dx~dy~g(x)~g(y)~T(x,y) + \cdots 
\label{S-matrix}
\ee
where $g$ is some test function. The expressions 
$T(x,y)$
are called {\it (second order) chronological products} because they must verify the causality property:
\be
T(x,y) = T(x) T(y)
\ee
for 
$x \succ y$
i.e. 
$(x - y)^{2} \geq 0, x^{0} - y^{0} \geq 0$;
in other words the point $x$ succeeds causally the point $y$. This is some generalization of the property
\be
U(t,s) = U(t,r) U(r,s),~~t > r > s
\ee
of the time evolution operator from non-relativistic quantum mechanics.

We go to the second order of perturbation theory using the {\it causal commutator}
\be
D^{A,B}(x,y) \equiv D(A(x),B(y)) = [ A(x),B(y)]
\ee
where 
$
A(x), B(y)
$
are arbitrary Wick monomials. These type of distributions are translation invariant i.e. they depend only on 
$
x - y
$
and the support is inside the light cones:
\be
supp(D) \subset V^{+} \cup V^{-}.
\ee

A theorem from distribution theory guarantees that one can causally split this distribution:
\be
D(A(x),B(y)) = A(A(x),B(y)) - R(A(x),B(y)).
\ee
where:
\be
supp(A) \subset V^{+} \qquad supp(R) \subset V^{-}.
\ee
The expressions 
$
A(A(x),B(y)), R(A(x),B(y))
$
are called {\it advanced} resp. {\it retarded} products. They are not uniquely defined: one can modify them with {\it quasi-local terms} i.e. terms proportional with
$
\delta(x - y)
$
and derivatives of it. 

There are some limitations on these redefinitions coming from Lorentz invariance and {\it power counting}: this means that we should not make the various distributions appearing in the advanced and retarded products too singular.

Then we define the {\it chronological product} by:
\be
T(A(x),B(y)) = A(A(x),B(y)) + B(y) A(x) = R(A(x),B(y)) + A(x) B(y).
\ee

The expression
$
T(x, y)
$
corresponds to the choice
\be
T(x,y) \equiv T(T(x), T(x)).
\ee

The ``naive'' definition
\be
T(A(x),B(y)) = \theta(x^{0}-y^{0}) A(x) B(y) + \theta(y^{0}-x^{0}) B(y) A(x)
\ee
involves an illegal operation, namely the multiplication of distributions. 
This appears in some loop contributions (the famous ultraviolet divergences). 

We will need in the following the causal commutator
\be
D^{IJ}(x,y) \equiv D(T^{I}(x), T^{J}(y)) = [T^{I}(x), T^{J}(y)]
\label{causal-comm-2}
\ee
where $[\cdot,\cdot]$ is always the graded commutator.
\newpage

\section{\bf The Yang-Mills Lagrangian\label{Lym}}

\subsection{First Order of the Perturbation Theory}

We consider the framework and notations from the end of Section \ref{q}. Then we have the following
result which describes the most general form of the Yang-Mills interaction.
Summation over the dummy indexes is used everywhere.

Let $T$ be a co-cycle for the operator$s$ which is as least tri-linear in the fields and is of canonical dimension
$
\omega(T) \leq 4
$
and ghost number
$
gh(T) = 0.
$
Then:
(i) $T$ is cohomologous to a non-trivial co-cycle of the form:
\bea
T = f_{abc} \left( \frac{1}{2}~v_{a\mu}~v_{b\nu}~F_{c}^{\nu\mu}
+ u_{a}~v_{b}^{\mu}~\partial_{\mu}\tilde{u}_{c}\right)
\nonumber \\
+ f^{\prime}_{abc} [\Phi_{a}~(\partial^{\mu}\phi_{b} - m~v_{b}^{\mu})~v_{c\mu} +
m_{b}~\Phi_{a}~\tilde{u}_{b}~u_{c} ]
\nonumber \\
+ \frac{1}{3!}~f^{\prime\prime}_{abc}~\Phi_{a}~\Phi_{b}~\Phi_{c}
+ \frac{1}{4!}~\sum_{a,b,c,d \in
I_{3}}~g_{abcd}~\Phi_{a}~\Phi_{b}~\Phi_{c}~\Phi_{d}
+ j^{\mu}_{a}~v_{a\mu} + j_{a}~\Phi_{a};
\label{T-sm}
\eea

(ii) The relation 
$
d_{Q}T = i~\partial_{\mu}T^{\mu}
$
is verified by:
\be
T^{\mu} = f_{abc} \left( u_{a}~v_{b\nu}~F^{\nu\mu}_{c} -
\frac{1}{2} u_{a}~u_{b}~\partial^{\mu}\tilde{u}_{c} \right)
+ f^{\prime}_{abc}~\Phi_{a}~\phi_{b}^{\mu}~u_{c}
+ j^{\mu}_{a}~u_{a}
\label{Tmu}
\ee

(iii) The relation 
$
d_{Q}T^{\mu} = i~\partial_{\nu}T^{\mu\nu}
$
is verified by:
\be
T^{\mu\nu} \equiv \frac{1}{2} f_{abc}~u_{a}~u_{b}~F_{c}^{\mu\nu}.
\ee

Here
\be
j^{\mu}_{a} \equiv \bar{\Psi}~t^{\epsilon}_{a} \otimes \gamma^{\mu}\gamma_{\epsilon}~\Psi, \qquad
j_{a} \equiv \bar{\Psi}~s^{\epsilon}_{a} \otimes \gamma_{\epsilon}~\Psi.
\ee
There are various restrictions on the constants appearing in the preceding 
expressions. We are interested in the structure of the coefficients
$
f_{abc}
$
and
$
f^{\prime}_{abc}
$
determining the electro-weak sector. We can imposed the following restrictions:

\be
f^{\prime}_{abc} = - (a \leftrightarrow b)
\label{c1}
\ee
\be
f^{\prime}_{abc} = 0,\qquad \forall c \in I_{3}
\label{c2}
\ee
\be
f^{\prime}_{abc} = 0,\qquad \forall a \in I_{1}
\label{c3}
\ee

The preceding expressions
$
T^{I}
$
are self-adjoint if the constants
$
f_{abc},~f^{\prime}_{abc}
$
are real. They also verify the following restrictions:
\be
f^{\prime}_{cab}~m_{a} - f^{\prime}_{cba}~m_{b}
= f_{abc}~m_{c},~\forall a,b \in I_{1} \cup I_{2},~c \in I_{2} \cup I_{3}
\ee
\be
m_{a}~s^{\epsilon}_{a} = M t^{\epsilon}_{a} - t^{- \epsilon}_{a} M
\label{mass1}
\ee
\be
[M, t_{a}] = 0, \quad  \{M, t'_{a}\} = 0, \quad \forall m_{a} = 0.
\label{mass2}
\ee
\be
f^{\prime\prime}_{abc} = \left\{\begin{array}{rcl} 
\frac{1}{m_{c}}~f'_{abc}~(m_{a}^{2} - m_{b}^{2}) & 
\mbox{for} & a, b \in I_{3}, c \in I_{2} \\
- \frac{1}{m_{c}}~f'_{abc}~m_{b}^{2} & 
\mbox{for} & a, c \in I_{2}, b \in I_{3}\\
0 & \mbox{for} & a, b, c \in I_{2}.\end{array}\right.
\label{f"}
\ee

Let us provide the proof in the simplest case when all spin $1$ fields are of null mass i.e. 
$
I_{2} = I_{3} = \emptyset.
$
We consider a Wick polynomial $T$ which
is tri-linear in the fields
$
v_{j}^{\mu}, u_{j}, \tilde{u}_{j}
$
has canonical dimension $4$ and null ghost number, is Lorentz covariant and gauge invariant in the sense (\ref{gauge}). 
First we list all possible monomials compatible with all these requirements; they are in the even sector (with respect to
parity):
\bea
T_{1} = f^{(1)}_{jkl} v_{j}^{\mu} v_{k}^{\nu} \partial_{\mu}v_{l\mu} 
\nonumber \\
T_{2} = f^{(2)}_{jkl} v_{j}^{\mu} v_{k\mu} \partial_{\nu}v_{l}^{\nu} \qquad
f^{(2)}_{jkl} = f^{(2)}_{kjl}
\eea
and
\be
T_{3} = g^{(1)}_{jkl} v^{\mu}_{j} u_{k} \partial_{\mu}\tilde{u}_{l} \qquad
T_{4} = g^{(2)}_{jkl} \partial_{\mu}v^{\mu}_{j} u_{k} \tilde{u}_{l} \qquad
T_{5} = g^{(3)}_{jkl} v^{\mu}_{j} \partial_{\mu}u_{k} \tilde{u}_{l}.
\ee 

We now list the possible trivial Lagrangians. They are obtained from total divergences of null ghost number
\bea
t^{(1)}_{\mu} = t^{(1)}_{jkl} v_{j}^{\nu} v_{k\nu} v_{l\mu} \qquad 
t^{(1)}_{jkl} = t^{(1)}_{kjl}
\nonumber \\
t^{(2)}_{\mu} = t^{(2)}_{jkl} v_{j\mu} u_{k} \tilde{u}_{l}
\eea 
and the co-boundary terms of ghost number $- 1$:
\bea
b^{(1)} = b^{(1)}_{jkl} v_{j}^{\mu} v_{k\mu} \tilde{u}_{l} \qquad 
b^{(1)}_{jkl} = b^{(1)}_{kjl}
\nonumber \\
b^{(2)} = b^{(2)}_{jkl} u_{j} \tilde{u}_{k} \tilde{u}_{l} \qquad 
b^{(2)}_{jkl} = - b^{(2)}_{jlk}.
\eea

Now we proceed as follows: using
$
\partial^{\mu}t^{(1)}_{\mu}
$
it is possible to make
\be
f^{(1)}_{jkl} = - f^{(1)}_{lkj};
\label{A1}
\ee
using 
$d_{Q}b^{(1)}$ 
we can make
\be
f^{(2)}_{jkl} = 0;
\ee
using
$
\partial^{\mu}t^{(2)}_{\mu}
$
it is possible to take
\be
g^{(3)}_{jkl} = 0;
\ee
finally, using 
$d_{Q}b^{(2)}$ 
we can make
\be
g^{(2)}_{jkl} = g^{(2)}_{kjl}.
\label{S1}
\ee

If we compute 
$
d_{Q}T
$
the result is
\be
d_{Q}T = i u_{j} A_{j}  + {\rm total~div}
\ee
where:
\bea
A_{j} = - 2 f^{(1)}_{jkl}~\partial^{\nu}v^{\mu}_{k}~\partial_{\mu}v_{l\nu}
+ (f^{(1)}_{lkj} + g^{(2)}_{kjl})~
\partial_{\mu}v^{\mu}_{k}~\partial_{\nu}v_{l}^{\nu}
\nonumber \\
+ (- f^{(1)}_{jkl} + f^{(1)}_{lkj} + f^{(1)}_{klj} + g^{(1)}_{kjl})~
v^{\mu}_{k}~\partial_{\mu}\partial_{\nu}v^{\nu}_{l}.
\eea

Now the gauge invariance condition (\ref{gau1}) becomes
\be
u_{j} A_{j} = \partial_{\mu}t^{\mu}.
\ee
From power counting arguments it follows that the general form for 
$
t^{\mu}
$
is
\be
t^{\mu} = u_{j} t^{\mu}_{j} + (\partial_{\nu}u_{j}) t^{\mu\nu}_{j}.
\ee
We can prove that
$
t^{\mu\nu}_{j} = g^{\mu\nu}~t_{j}
$
from where
$
A_{j} = - \partial^{2}t_{j}.
$
Making a general ansatz for 
$
t_{j}
$
we obtain that we must have in fact 
\be
A_{j} = 0
\ee
i.e. the following system of equations:
\bea
f^{(1)}_{jkl} = - f^{(1)}_{kjl}
\nonumber \\
f^{(1)}_{lkj} + g^{(2)}_{kjl} = 0
\nonumber \\
- f^{(1)}_{jkl} + f^{(1)}_{lkj} + f^{(1)}_{klj} + g^{(1)}_{kjl} = 0.
\label{A2}
\eea
The first equation, together with (\ref{A1}) amounts to the total antisymmetry of the expression
$
f_{jkl} \equiv f^{(1)}_{jkl};
$
the second equation from the preceding system gives then
$
g^{(2)}_{kjl} = 0.
$
The odd sector (with respect to parity) does not give non-trivial solutions so we obtain the (unique) solution:
\be
T = f^{(1)}_{jkl} ( v_{j}^{\mu} v_{k}^{\nu} \partial_{\nu}v_{l\mu}
- v_{j}^{\mu} u_{k} \partial_{\mu}\tilde{u}_{l});
\label{tym}
\ee
it can be easily be proved that 
$
d_{Q}t^{(1)}
$
is indeed a total divergence.

\newpage

\subsection{Second Order Gauge Invariance\label{second}}

Second order gauge invariance is best treated in the off-shell formalism \cite{caciulata4}. It essentially mens to construct the 
Hilbert space as in the preceding Section but to replace everywhere
$
D_{m}
$
by some off-shell distribution 
$
D_{m}^{\rm off}
$
which does not verify Klein-Gordon equation but converges in some limit (in the sense of distribution theory) to
$
D_{m}.
$ 
For instance we can take 
\be
D_{m,\epsilon}^{\rm off} \equiv \int d\lambda \rho_{m,\epsilon}(\lambda) D_{\lambda}
\ee
where 
$
\rho_{m,\epsilon}(\lambda)
$
is a function depending some parameter $\epsilon$ and converging, say for 
$
\epsilon \rightarrow 0
$
to the distribution 
$
\delta(\lambda - m).
$

One computes the second order causal commutator(\ref{causal-comm-2}) and finds out that the tree contribution has the following generic form:
\be
D^{IJ}(x,y)^{\rm tree} = [T^{I}(x),T^{J}(y)]^{\rm tree} = 
\sum_{m}~[~D_{m}(x - y)~A^{IJ}_{m}(x,y) 
+ \partial_{\rho}~D_{m}(x - y)~A^{IJ;\rho}_{m}(x,y) ]
\label{T-tree} 
\ee
where the sum runs over the various masses from the spectrum of the model and the expressions
$
A^{IJ}_{m}
$
and
$
A^{IJ;\rho}_{m}
$
are Wick polynomials. Then we apply the BRST operator to this commutator and obtain
\be
sD^{IJ}(x,y)^{\rm tree}(x,y) = 
\sum_{m}~[~K_{m}D_{m}(x - y)~A^{IJ}_{m}(x,y) 
+ \partial_{\rho}~K_{m}D_{m}(x - y)~A^{IJ;\rho}_{m}(x,y) ]
\label{sD}
\ee
where 
$K_{m}$
is the Klein-Gordon operator. Obviously, we get zero in the on-shell limit. 

Next, we observe that we can take
\bea
T^{IJ}(x,y)^{\rm tree} = 
\sum_{m}~[~D_{m}^{F}(x - y)~A^{IJ}_{m}(x,y) 
+ \partial_{\rho}~D_{m}^{F}(x - y)~A^{IJ;\rho}_{m}(x,y) ]
\eea
and all Bogoliubov axioms are true (in the second order) except gauge invariance. 
Indeed, from (\ref{sD}) we obtain a similar relation with
$
D_{m} \rightarrow D_{m}^{F}:
$
\be
sT^{IJ}(x,y)^{\rm tree}(x,y) = 
\sum_{m}~[~K_{m}D^{F}_{m}(x - y)~A^{IJ}_{m}(x,y) 
+ \partial_{\rho}~K^{F}_{m}D_{m}(x - y)~A^{IJ;\rho}_{m}(x,y) ]
\label{sT}
\ee
so in the  on-shell limit we get $\delta$ terms (anomalies) due to
\be
K_{m}D^{F}_{m}(x-y) = \delta(x-y).
\ee

\newpage
We list below the result of the off-shell computations for the formula (\ref{sD}), 
which are rather involved in the general case; 
the case of QCD was treated in \cite{caciulata4}. For simplicity we denote
$
K_{e} \equiv K_{m_{e}}
$
(in the Yang-Mills sector) and
$
K_{C} \equiv K_{m_{C}}
$ 
in the Dirac sector. 
We first have 
\be
sD^{IJ}(x,y) = [ sT^{I}(x), T^{J}(y) ] - (-1)^{|I||J|}~[sT^{J}(y), T^{I}(x) ]
\label{sD1}
\ee
so we need to compute only the (graded) commutator:
\be
S^{IJ}(x,y) \equiv [ sT^{I}(x), T^{J}(y) ] 
\label{SIJ}
\ee

It is useful to list the non-null expressions
\be
S^{I} \equiv sT^{I};
\ee
they are:
\be
S = S^{\emptyset} = \sum_{j=1}^{5}~S_{j}
\ee
where
\bea
S_{1} \equiv i~f_{abc}~u_{a}~v_{b}^{\mu}~K_{c}v_{c\mu}
\nonumber \\ 
S_{2} \equiv \frac{i}{2}~f_{abc}~u_{a}~u_{b}~K_{c}\tilde{u}_{c}
\nonumber \\
S_{3} \equiv - i~f^{\prime}_{bca}~u_{a}~\Phi_{b}~K_{c}\Phi_{c}
\nonumber \\
S_{4} \equiv 
- i~u_{a}~\partial_{\mu}\bar{\Psi}~t_{a}^{\epsilon} \otimes \gamma^{\mu}~\gamma_{\epsilon}~\Psi
- u_{a}~\bar{\Psi}~M~t_{a}^{\epsilon} \otimes \gamma_{\epsilon}~\Psi
\nonumber \\
S_{5} \equiv 
- i~u_{a}~\bar{\Psi}~t_{a}^{\epsilon} \otimes \gamma^{\mu}~\gamma_{\epsilon}~\partial_{\mu}\Psi
+ u_{a}~\bar{\Psi}~t_{a}^{- \epsilon}~M \otimes \gamma_{\epsilon}~\Psi
\nonumber \\
= - i~u_{a}~\bar{\Psi}~t_{a}^{-\epsilon} \otimes \gamma_{\epsilon}~\gamma^{\mu}\partial_{\mu}\Psi
+ u_{a}~\bar{\Psi}~t_{a}^{- \epsilon}~M \otimes \gamma_{\epsilon}~\Psi
\eea
and
\be
S^{\mu} \equiv \frac{i}{2}~f_{abc}~u_{a}~u_{b}~K_{c}v_{c}^{\mu}
\ee

We can proceed to the computation of the expressions
$
S^{IJ}.
$
We have the following non-trivial cases:
\vskip 0.5cm
\centerline{ 1. Case $I = J = \emptyset$}

\bea
S^{\emptyset\emptyset}(x,y) = \sum_{e}~K_{e}(x - y)~A^{\emptyset\emptyset}_{e}(x,y) 
+ \partial_{\mu}K_{e}(x - y)~A^{\emptyset\emptyset;\mu}_{e}(x,y)
\nonumber\\
+ \sum_{C}~K_{C}(x - y)~F^{\emptyset\emptyset}_{C}(x,y) + \cdots
\eea
where $\cdots$ are terms null on-shell (i.e. involving the equations of motion) and
\bea
A^{\emptyset\emptyset}_{e}(x,y) = f_{abe}~f_{cde}~[- u_{a}(x)~v_{b\mu}(x)~v_{c\nu}(y)~F_{d}^{\nu\mu}(y)
+ u_{a}(x)~v_{b}^{\mu}(x)~u_{c}(y)~\partial_{\mu}\tilde{u}_{d}(y)
\nonumber\\
- \frac{1}{2}~u_{a}(x)~u_{b}(x)~v_{c}^{\mu}(y)~\partial_{\mu}\tilde{u}_{d}(y) ]
\nonumber\\
+ f_{abe}~f^{\prime}_{cde}~[- u_{a}(x)~v_{b}^{\mu}(x)~\Phi_{c}(y)~\partial_{\mu}\Phi_{d}(y)
+ m_{d}~u_{a}(x)~v_{b}^{\mu}(x)~\Phi_{c}(y)~v_{d\mu}(y)
\nonumber\\
+ \frac{1}{2}~m_{d}~u_{a}(x)~u_{b}(x)~\Phi_{c}(y)~\tilde{u}_{d}(y) ]
\nonumber \\
- f_{abe}~f^{\prime}_{ecd}~m_{e}~u_{a}(x)~v_{b}^{\mu}(x)~\Phi_{c}(y)~v_{d\mu}(y)
- f_{abe}~u_{a}(x)~v_{b}^{\mu}(x)~j_{e\mu}(y)
\nonumber\\
+ f^{\prime}_{eba}~f^{\prime}_{ecd}~[ u_{a}(x)~\Phi_{b}(x)~\partial_{\mu}\Phi_{c}(y)~v_{d}^{\mu}(y)
- m_{c}~u_{a}(x)~\Phi_{b}(x)~v_{c}^{\mu}(y)~v_{d\mu}(y)
\nonumber\\
+ m_{c}~u_{a}(x)~\Phi_{b}(x)~\tilde{u}_{c}(y)~u_{d}(y) ]
\nonumber\\
+ \frac{1}{2}~m_{d}~f^{\prime}_{eba}~f^{\prime\prime}_{ecd}~u_{a}(x)~\Phi_{b}(x)~\Phi_{c}(y)~\Phi_{d}(y)
+ f^{\prime}_{eba}~u_{a}(x)~\Phi_{b}(x)~j_{c}(y)
\label{A}
\eea
\be
A^{\emptyset\emptyset;\mu}_{e}(x,y) = f_{abe}~f_{cde}~u_{a}(x)~v_{b}^{\rho}(x)~v_{c\rho}(y)~v_{d}^{\mu}(y)
+ f^{\prime}_{eba}~f^{\prime}_{ecd}~u_{a}(x)~\Phi_{b}(x)~\Phi_{c}(y)~v_{d}^{\mu}(y)
\ee
\bea
F^{\emptyset\emptyset}_{C}(x,y) = - i~(t^{\epsilon}_{a})_{AC}~(t^{\epsilon}_{b})_{CB}~
u_{a}(x)~v_{b}^{\mu}(y)~\bar{\Psi}_{A}(x)~\gamma_{\mu}~\gamma_{\epsilon}~\Psi_{B}(y)
\nonumber\\
+ i~(t^{\epsilon}_{b})_{AC}~(t^{\epsilon}_{a})_{CB}~
u_{a}(x)~v_{b}^{\mu}(y)~\bar{\Psi}_{A}(y)~\gamma_{\mu}~\gamma_{\epsilon}~\Psi_{B}(x)
\nonumber\\
- i~(t^{- \epsilon}_{a})_{AC}~(s^{\epsilon}_{b})_{CB}~
u_{a}(x)~\Phi_{b}(y)~\bar{\Psi}_{A}(x)~\gamma_{\epsilon}~\Psi_{B}(y)
\nonumber\\
+ i~(s^{\epsilon}_{b})_{AC}~(t^{\epsilon}_{a})_{CB}~
u_{a}(x)~\Phi_{b}(y)~\bar{\Psi}_{A}(y)~\gamma_{\epsilon}~\Psi_{B}(x)
\eea

\centerline{2. Case $I = [\mu], J = \emptyset $ }
\vskip 0.5cm
\be
S^{[\mu]\emptyset}(x,y) = \sum_{e}~K_{e}(x - y)~A^{[\mu] \emptyset}_{e}(x,y) 
+ \partial_{\nu}K_{e}(x - y)~A^{[\mu] \emptyset;\nu}_{e}(x,y) + \cdots
\ee
where
\bea
A^{[\mu] \emptyset}_{e}(x,y) = f_{abe}~f_{cde}~[- u_{a}(x)~u_{b}(x)~v_{c\nu}(y)~F_{d}^{\nu\mu}(y)
+ u_{a}(x)~u_{b}(x)~u_{c}(y)~\partial^{\mu}\tilde{u}_{d}(y) ]
\nonumber\\
+ \frac{1}{2}~f_{abe}~f^{\prime}_{cde}~[- u_{a}(x)~u_{b}(x)~\Phi_{c}(y)~\partial^{\mu}\Phi_{d}(y)
+ m_{d}~u_{a}(x)~u_{b}(x)~\Phi_{c}(y)~v_{d}^{\mu}(y) ]
\nonumber \\
- \frac{1}{2}~f_{abe}~f^{\prime}_{ecd}~m_{e}~u_{a}(x)~u_{b}(x)~\Phi_{c}(y)~v_{d}^{\mu}(y)
%\nonumber\\
- \frac{1}{2}~f_{abe}~u_{a}(x)~u_{b}(x)~j_{e}^{\mu}(y)
\eea
\be
A^{[\mu] \emptyset;\nu}_{e}(x,y) = \frac{1}{2}~f_{abe}~f_{cde}~u_{a}(x)~u_{b}(x)~v_{c}^{\mu}(y)~v_{d}^{\nu}(y)
\ee
\vskip 0.5cm
\centerline{ 3. Case $I = \emptyset, J = [\mu]$}

\bea
S^{\emptyset [\mu]}(x,y) = \sum_{e}~K_{e}(x - y)~A^{\emptyset [\mu]}_{e}(x,y) 
+ \partial_{\nu}K_{e}(x - y)~A^{\emptyset [\mu];\nu}_{e}(x,y)
\nonumber\\
+ \sum_{C}~K_{C}(x - y)~F^{\emptyset [\mu]}_{C}(x,y) + \cdots
\eea
where:
\bea
A^{\emptyset [\mu]}_{e}(x,y) = f_{abe}~f_{cde}~[ u_{a}(x)~v_{b\nu}(x)~u_{c}(y)~F_{d}^{\nu\mu}(y)
- \frac{1}{2}~u_{a}(x)~u_{b}(x)~v_{c\nu}(y)~F_{d}^{\nu\mu}(y)
\nonumber\\
+ \frac{1}{2}~u_{a}(x)~u_{b}(x)~u_{c}(y)~\partial^{\mu}\tilde{u}_{d}(y)]
\nonumber\\
- f_{abe}~f^{\prime}_{ecd}~m_{e}~u_{a}(x)~v_{b}^{\mu}(x)~\Phi_{c}(y)~u_{d}(y)
\nonumber\\
- \frac{1}{2}~f_{abe}~f^{\prime}_{cde}~[ u_{a}(x)~u_{b}(x)~\Phi_{c}(y)~\partial^{\mu}\Phi_{d}(y)
- m_{d}~u_{a}(x)~u_{b}(x)~\Phi_{c}(y)~v_{d}^{\mu}(y) ]
\nonumber \\
- \frac{1}{2}~f_{abe}~u_{a}(x)~u_{b}(x)~j_{e}^{\mu}(y)
\nonumber\\
+ f^{\prime}_{eba}~f^{\prime}_{ecd}~[ u_{a}(x)~\Phi_{b}(x)~\partial_{\mu}\Phi_{c}(y)~u_{d}(y)
- m_{c}~u_{a}(x)~\Phi_{b}(x)~v_{c}^{\mu}(y)~u_{d}(y) ]
\eea
\bea
A^{\emptyset [\mu];\nu}_{e}(x,y) = f_{abe}~f_{cde}~[ u_{a}(x)~v_{b}^{\mu}(x)~u_{c}(y)~v_{d}^{\nu}(y)
- \eta^{\mu\nu}~u_{a}(x)~v_{b}^{\rho}(x)~u_{c}(y)~v_{d\rho}(y) ]
\nonumber\\
+ f^{\prime}_{eba}~f^{\prime}_{ecd}~\eta^{\mu\nu}~u_{a}(x)~\Phi_{b}(x)~\Phi_{c}(y)~u_{d}(y)
\eea
\bea
F^{\emptyset [\mu]}_{C}(x,y) = - i~(t^{\epsilon}_{a})_{AC}~(t^{\epsilon}_{b})_{CB}~
u_{a}(x)~u_{b}(y)~\bar{\Psi}_{A}(x)~\gamma^{\mu}~\gamma_{\epsilon}~\Psi_{B}(y)
\nonumber\\
+ i~(t^{\epsilon}_{b})_{AC}~(t^{\epsilon}_{a})_{CB}~
u_{a}(x)~u_{b}(y)~\bar{\Psi}_{A}(y)~\gamma^{\mu}~\gamma_{\epsilon}~\Psi_{B}(x)
\eea

\vskip 0.5cm
\centerline{ 4. Case $I = [\mu], J = [\nu]$}

\be
S^{[\mu] [\nu]}(x,y) = \sum_{e}~K_{e}(x - y)~A^{[\mu] [\nu]}_{e}(x,y) 
+ \partial_{\rho}K_{e}(x - y)~A^{[\mu] [\nu];\rho}_{e}(x,y) + \cdots
\ee
where:
\bea
A^{[\mu] [\nu]}_{e}(x,y) = \frac{1}{2}~f_{abe}~f_{cde}~u_{a}(x)~u_{b}(x)~u_{c}(y)~F_{d}^{\mu\nu}(y)
\nonumber\\
- \frac{1}{2}~f_{eab}~f^{\prime}_{ecd}~m_{e}~\eta^{\mu\nu}~u_{a}(x)~u_{b}(x)~\Phi_{c}(y)~u_{d}(y)
\eea
\be
A^{[\mu],[\nu];\rho}_{e}(x,y) = \frac{1}{2}~f_{abe}~f_{cde}~[ \eta^{\mu\nu}~u_{a}(x)~u_{b}(x)~u_{c}(y)~v_{d}^{\rho}(y)
- \eta^{\nu\rho}~u_{a}(x)~u_{b}(x)~u_{c}(y)~v_{d}^{\mu}(y) ]
\ee

\vskip 0.5cm
\centerline{ 5. Case $I = \emptyset, J = [\mu\nu]$}

\be
S^{\emptyset [\mu\nu]}(x,y) = \sum_{e}~K_{e}(x - y)~A^{\emptyset [\mu\nu]}_{e}(x,y) 
+ \partial_{\rho}K_{e}(x - y)~A^{\emptyset [\mu\nu];\rho}_{e}(x,y) + \cdots
\ee
where:
\be
A^{\emptyset [\mu\nu]}_{e}(x,y) = - \frac{1}{2}~f_{abe}~f_{cde}~u_{a}(x)~u_{b}(x)~u_{c}(y)~F_{d}^{\mu\nu}(y)
\ee
\be
A^{\emptyset [\mu\nu];\rho}_{e}(x,y) = \frac{1}{2}~f_{abe}~f_{cde}~[ \eta^{\mu\rho}~u_{a}(x)~v_{b}^{\nu}(x)~u_{c}(y)~u_{d}(y)
- (\mu \leftrightarrow \nu) ]
\ee

\vskip 0.5cm
\centerline{ 6. Case $I = [\rho], J = [\mu\nu]$}

\be
S^{[\rho] [\mu\nu]}(x,y) = \partial_{\sigma}K_{e}(x - y)~A^{[\rho] [\mu\nu];\sigma}_{e}(x,y) + \cdots
\ee
where:
\be
A^{[\rho] [\mu\nu];\sigma}_{e}(x,y) = - \frac{1}{4}~f_{abe}~f_{cde}~
( \eta^{\mu\rho}~\eta^{\nu\sigma} - \eta^{\nu\rho}~\eta^{\mu\sigma} )~
u_{a}(x)~u_{b}(x)~u_{c}(y)~u_{d}(y)
\ee

Now, from (\ref{sD1}) and (\ref{SIJ}) we get 
\bea
sD^{IJ}(x,y)^{\rm tree} = \sum_{e}~K_{e}D_{e}(x - y)~W^{IJ}_{e}(x,y) + 
\sum_{e}~\partial_{\rho}K_{e}D_{e}(x - y)~W^{IJ;\rho}_{e}(x,y)
\nonumber\\
+ \sum_{e}K_{C}D_{C}(x - y)~V^{IJ}_{C}(x,y) + 
\eea
where
\bea
W^{IJ}_{e}(x,y)\equiv A^{IJ}_{e}(x,y) + (-1)^{|I||J|}~A^{JI}_{e}(y,x)
\nonumber\\
W^{IJ;\rho}_{e}(x,y)\equiv A^{IJ;\rho}_{e}(x,y) - (-1)^{|I||J|}~A^{JI;\rho}_{e}(y,x)
\nonumber\\
V^{IJ}_{C}(x,y)\equiv F^{IJ}_{C}(x,y) + (-1)^{|I||J|}~F^{JI}_{C}(y,x).
\eea

The next combinatorial step is to eliminate in a systematic way the derivatives on 
$
K_{m}
$;
this can be done, defining the {\it renormalized off-shell} causal commutators:
\be
D^{IJ}_{\rm ren}(x,y) \equiv D^{IJ}(x,y) + \sum_{e}K_{e}D_{e}(x - y)~C^{IJ}_{e}(x,y)
\label{D-ren}
\ee
where the non-zero Wick polynomials
$
C^{IJ}_{e}
$
are given by the following formulas
\be
C^{[\mu\nu][\rho\sigma]}_{e}(x,y) \equiv \frac{i}{4}~f_{abe}~f_{cde}~( \eta^{\mu\rho}~\eta^{\nu\sigma} - \eta^{\nu\rho}~\eta^{\mu\sigma} )~
u_{a}(x)~u_{b}(x)~u_{c}(y)~u_{d}(y)
\ee
\be
C^{[\mu\nu][\rho]}_{e}(x,y) \equiv - \frac{i}{2}~f_{abe}~f_{cde}~[ \eta^{\mu\rho}~u_{a}(x)~u_{b}(x)~u_{c}(y)~v^{\nu}_{d}(y)
- ( \mu \leftrightarrow \nu ) ]
\ee
\be
C^{[\mu\nu]\emptyset}_{e}(x,y) \equiv - \frac{i}{2}~f_{abe}~f_{cde}~u_{a}(x)~u_{b}(x)~v^{\mu}_{c}(y)~v^{\nu}_{d}(y)
\ee
\bea
C^{[\mu][\nu]}_{e}(x,y) \equiv - i~f_{abe}~f_{cde}~[ u_{a}(x)~v^{\nu}_{b}(x)~u_{c}(y)~v^{\mu}_{d}(y)
- \eta^{\mu\nu}~u_{a}(x)~v^{\rho}_{b}(x)~u_{c}(y)~v_{d\rho}(y) ]
\nonumber\\
- i~f^{\prime}_{eba}~f^{\prime}_{edc}~\eta^{\mu\nu}~u_{a}(x)~\Phi_{b}(x)~u_{c}(y)~\Phi_{d}(y)
\eea
\be
C^{[\mu]\emptyset}_{e}(x,y) \equiv - i~f_{abe}~f_{cde}~u_{a}(x)~v^{\nu}_{b}(x)~v_{c\nu}(y)~v^{\mu}_{d}(y)
- i~f^{\prime}_{eba}~f^{\prime}_{edc}~\eta^{\mu\nu}~u_{a}(x)~\Phi_{b}(x)~v^{\mu}_{c}(y)~\Phi_{d}(y)
\ee
\be
C^{\emptyset\emptyset}_{e}(x,y) \equiv \frac{i}{2}~f_{abe}~f_{cde}~v^{\mu}_{a}(x)~v^{\nu}_{b}(x)~v_{c\mu}(y)~v_{d\nu}(y)
- i~f^{\prime}_{eba}~f^{\prime}_{edc}~v_{a\mu}(x)~\Phi_{b}(x)~v^{\mu}_{c}(y)~\Phi_{d}(y)
%\nonumber\\
%+  \frac{i}{2}~\sum_{e \in I_{2}}~f^{\prime}_{abe}~f^{\prime\prime}_{cde}~[ \Phi_{a}(x)~\Phi_{b}(x)~\Phi_{c}(y)~\Phi_{d}(y)
%+ ( x \leftrightarrow y).
\ee

Then a tedious but straightforward computation gives
\be
sD^{IJ}_{\rm ren}(x,y) = \sum_{e}K_{e}D_{e}(x - y)~w^{IJ}_{e}(x,y) + \cdots
\label{sDren}
\ee
where the non-zero Wick monomials
$
a^{IJ}_{e}
$
are:
\be
w^{[\mu\nu]\emptyset}_{e}(x,y) = A^{\emptyset [\mu\nu]}_{e}(x,y) 
- \frac{1}{2}~f_{abe}~f_{cde}~u_{a}(x)~u_{b}(x)~u_{c}(y)~F^{\mu\nu}_{d}(y)
\ee
\be
w^{[\mu][\nu]}_{e}(x,y) = A^{[\mu][\nu]}_{e}(x,y) - ( x \leftrightarrow y, \mu \leftrightarrow \nu)
\ee
\bea
w^{[\mu]\emptyset}_{e}(x,y) = A^{[\mu]\emptyset}_{e}(x,y) + A^{\emptyset[\mu]}_{e}(y,x)
- f_{abe}~f_{cde}~u_{a}(x)~v_{b\nu}(x)~u_{c}(y)~F^{\mu\nu}_{d}(y)
\nonumber\\
+ f^{\prime}_{eba}~f^{\prime}_{edc}~[ u_{a}(x)~\Phi_{b}(x)~u_{c}(y)~\partial^{\mu}\Phi_{d}(y)
\nonumber\\
+ m_{a}~u_{a}(x)~u_{b}(x)~v^{\mu}_{c}(y)~\Phi_{d}(y) - m_{c}~u_{a}(x)~\Phi_{b}(x)~v^{\mu}_{c}(y)~u_{d}(y) ]
\eea
\bea
w^{\emptyset\emptyset}_{e}(x,y) = \{ A^{\emptyset\emptyset}_{e}(x,y)
+ \frac{1}{2}~f_{abe}~f_{cde}~u_{a}(x)~F_{b\mu\nu}(x)~v^{\mu}_{c}(y)~v^{\nu}_{d}(y)
\nonumber\\
- f^{\prime}_{eba}~f^{\prime}_{edc}~[ u_{a}(x)~\partial_{\mu}\Phi_{b}(x)~v^{\mu}_{c}(y)~\Phi_{d}(y)
%\nonumber\\
- m_{a}~v_{a\mu}(x)~u_{b}(x)~v^{\mu}_{c}(y)~\Phi_{d}(y) ] \} + (x \leftrightarrow y)
\eea

Now we construct the renormalized chronological products as in (\ref{D-ren})
\be
T^{IJ}_{\rm ren}(x,y) \equiv T^{IJ}(x,y) + \sum_{e}K^{F}_{e}D_{e}(x - y)~C^{IJ}_{e}(x,y)
\label{T-ren}
\ee
and obtain, similarly to the formula (\ref{sDren}
\be
sT^{IJ}_{\rm ren}(x,y) = \sum_{e}K_{e}D^{F}_{e}(x - y)~w^{IJ}_{e}(x,y) + \cdots
\label{sTren}
\ee

In the on-shell limit we get from above
\be
T^{IJ}(x,y) \rightarrow t^{IJ}(x,y),\qquad
T^{IJ}_{\rm ren}(x,y) \rightarrow t^{IJ}_{\rm ren}(x,y) 
\ee
and we get
\be
t^{IJ}_{\rm ren}(x,y) = t^{IJ}(x,y) + \delta (x - y)~N^{IJ}(x)
\label{T-ren-on-shell}
\ee
where
\be
N^{IJ}(x) = \sum_{e}~C^{IJ}(x,x).
\ee

Moreover we have
\be
st^{IJ}_{\rm ren}(x,y) = \delta (x - y)~w^{IJ}(x)
\label{sTren-on-shell}
\ee
where
\be
w^{IJ}(x) \equiv \sum_{e}~w^{IJ}_{e}(x,x) + \sum_{C}~V^{IJ}_{C}(x,x)
\ee
and the expression from the right hand side of (\ref{sTren-on-shell}) is an anomaly. The only way to
save gauge invariance is that the anomaly
\be
{\cal A}^{IJ}(x,y) \equiv \delta (x - y)~w^{IJ}(x)
\ee
is a co-boundary i.e. of the form
\be
sb^{IJ}(x,y) = (d_{Q} - i \delta)b^{IJ}(x,y)
\ee
with 
$
b^{IJ}
$
also quasi-local expressions:
\be
b^{IJ}(x,y) = \delta (x - y)~B^{I,J}(x)
\ee
with 
$
B^{IJ}(x)
$
some Wick polynomials with appropriate symmetry properties. Then an easy computation prove that this relation is true {\it iff}
\be
w^{IJ} = 0.
\ee
%\newpage
The preceding relation leads to:

\be
\sum_{c}~(f_{abc}~f_{dec} + f_{bdc}~f_{aec} + f_{dac}~f_{bec}) = 0
\label{jacoby}
\ee
(which is the Jacobi identity)
\be
\sum_{c}~[ f^{\prime}_{dca}~f^{\prime}_{ceb}
- (a \leftrightarrow b) ] = 
- \sum_{c}~f_{abc}~f^{\prime}_{dec},
\qquad
a,b \in I_{1} \cup I_{2},~d,e \in I_{2} \cup I_{3}.
\label{f21}
\ee
\be
[ t_{a}^{\epsilon}, t_{b}^{\epsilon} ] = i~f_{abc}~t_{c}^{\epsilon}
\label{representation}
\ee
\be
t_{a}^{- \epsilon}~s_{b}^{\epsilon} - s_{b}^{\epsilon}~t_{a}^{\epsilon}
= i~f^{\prime}_{bca}~s_{c}^{\epsilon} 
\label{tensor}
\ee
\be
S_{bcd}(f^{\prime}_{eba}~f^{\prime\prime}_{ecd}) = 0,~\forall a \in I_{1}
\ee
and the expression
\be
f_{a\{ bcd\}} \equiv \frac{1}{m_{a}}~S_{bcd}(f^{\prime}_{eba}~f^{\prime\prime}_{ecd}),~\forall a \in I_{2}
\ee
is of the form
\be
f_{a\{bcd\}} = F_{\{ abcd\},} ~\forall a \in I_{2}
\ee
where
$
F_{\{ abcd\}}
$
is completely symmetric.
\newpage

\newpage
\section{The Standard Model\label{sm}}

We consider the following particular case relevant for the electro-weak
sector of the standard model. The Lie algebra is real and isomorphic to
$
u(1) \times su(2)
$
and we have
$
I_{1} = \{0\}, I_{2} = \{1,2,3\}.
$
The non-zero constants
$
f_{abc}
$
are:
\be
f_{210} = \sin\theta, \quad f_{321} = \cos\theta
%\quad f_{310} = 0, \quad f_{320} = 0
\ee
with
$\cos \theta > 0$
and the other constants determined through the anti-symmetry property;
$
\theta
$
is the {\it Weinberg angle}. It is interesting to see that for a
four-dimensional
Lie algebra, the Jacobi identity is trivially verified. So there are two cases:
only one of the structure constants
$
f_{012}, f_{023}, f_{031}
$
is non-zero (and we end up with the case above after some re-scalings) and the 
case when at least two of the preceding structure constants are non-zero. The
last case leads to the equality of all masses and it is not interesting from the
physical point of view.

From the relation expressing first order gauge invariance of the preceding Section we obtain relatively easy:
\be
m_{1} = m_{2}
\label{m1}
\ee
and
\bea
f^{\prime}_{231} = f^{\prime}_{312} = - \cos\theta~\frac{m_{3}}{2 m_{1}}
\nonumber \\
f^{\prime}_{123} = - \cos\theta~\left( 1 - \frac{m_{3}^{2}}{2 m_{1}^{2}} \right)
\nonumber\\
f^{\prime}_{120} = - \sin\theta
\label{f12-1}
\eea
\bea
f^{\prime}_{121} = f^{\prime}_{131} = f^{\prime}_{122} = f^{\prime}_{232} =
f^{\prime}_{133} =
f^{\prime}_{233} = f^{\prime}_{130} = f^{\prime}_{230} = 0
\nonumber \\
f^{\prime}_{0bc} = 0, \forall b, c 
\nonumber \\
f^{\prime}_{j10} = f^{\prime}_{j20} = f^{\prime}_{j30} = 0,~\forall j \in I_{3} 
\label{f12-2}
\eea
\bea
f^{\prime}_{jab} = m_{a}~g_{jab},
\nonumber\\
g_{jab} = g_{jba}, \qquad \forall j \in I_{3}, \forall a, b = 1,2,3.
\label{f12-3}
\eea

The relation expressing second order gauge invariance are much harder to analyze. 

It all depends on the value of the parameter
$
\gamma = \frac{m_{3}~\cos\theta}{m_{1}}
$
defined in the Introduction. The computations are presented in detail in \cite{higgs}.

The first case is
$
\gamma = 1
$
and corresponds then to the usual standard model. We work out the second line of the interaction Lagrangian
(\ref{T-sm}). Because we have only one Higgs field i.e.
$
|I_{3}| = 1
$
we can take
$
I_{3} = \{ H \}
$
and the non-zero expressions
$
f^{\prime}_{abc}
$
from the second line of the interaction Lagrangian (\ref{T-sm}) are:
\bea
f^{\prime}_{321} = - f'_{312} = \frac{1}{2}, \quad
f^{\prime}_{123} = - \frac{\cos~2\theta}{2\cos~\theta},\quad
f^{\prime}_{210} = \sin~\theta
\nonumber\\
f^{\prime}_{H11} = f'_{H22} = \frac{1}{2}, \quad
f^{\prime}_{H33} = \frac{1}{2\cos~\theta}
\eea
and those following from the antisymmetry property in the first two indexes.
As a result we have the scalar + Yang-Mills interaction: 
\bea
T_{s + YM} = 
\sin\theta [ ( \Phi_{2}~\phi_{1\mu} - \Phi_{1}~\phi_{2\mu})~v_{0}^{\mu} 
+ m_{1}~( \Phi_{2}~\tilde{u}_{1} - \Phi_{1}~\tilde{u}_{2})~u_{0}
\nonumber\\
+ \frac{1}{2}~[ ( \Phi_{3}~\phi_{2\mu} - \Phi_{2}~\phi_{3\mu})~v_{1}^{\mu} 
+ (m_{2}~\Phi_{3}~\tilde{u}_{2} - m_{3} \Phi_{2}~\tilde{u}_{3})~u_{1}
\nonumber\\
+ ( \Phi_{1}~\phi_{3\mu} - \Phi_{3}~\phi_{1\mu})~v_{2}^{\mu} 
+ (m_{3}~\Phi_{1}~\tilde{u}_{3} - m_{1} \Phi_{3}~\tilde{u}_{1})~u_{2} ]
\nonumber\\
+ \frac{\cos2\theta}{2\cos\theta}~( \Phi_{2}~\phi_{1\mu} - \Phi_{1}~\phi_{2\mu})~v_{3}^{\mu} 
+ m_{1}~( \Phi_{2}~\tilde{u}_{1} - \Phi_{1}~\tilde{u}_{2})~u_{3}
\nonumber\\
+ \frac{1}{2}~[ ( \Phi_{H}~\phi_{1\mu} - \Phi_{1}~\partial_{\mu}\Phi_{H})~v_{1}^{\mu} 
+ m_{1}~\Phi_{H}~\tilde{u}_{1}~u_{1}
\nonumber\\
+ ( \Phi_{H}~\phi_{2\mu} - \Phi_{2}~\partial_{\mu}\Phi_{H})~v_{2}^{\mu} 
+ m_{1}~\Phi_{H}~\tilde{u}_{2}~u_{2} ]
\nonumber\\
+\frac{1}{2\cos\theta}~[ ( \Phi_{H}~\phi_{3\mu} - \Phi_{3}~\partial_{\mu}\Phi_{H})~v_{3}^{\mu} 
+ m_{3}~\Phi_{H}~\tilde{u}_{3}~u_{3} ].
\label{LagrangeanSM}
\eea

In the second case corresponding to
$
\gamma > 1
$
we can take
$
I_{3} = \{ H, H_{2}, K_{2}, \dots, H_{N}, K_{N} \}
$
so beside the Higgs field 
$
\Phi_{H}
$
there are some other (real) scalar fields
$
\Phi_{H_{n}},\Phi_{K_{n}},\quad n = 2,\dots,N.
$

The non-zero expressions
$
f^{\prime}_{abc}
$
from the second line of the interaction Lagrangian (\ref{T-sm}) are:
\bea
f^{\prime}_{321} = - f'_{312} = \frac{\gamma}{2}, \quad
f^{\prime}_{123} = - \frac{2\cos^{2}\theta - \gamma^{2}}{2cos~\theta},\quad
f^{\prime}_{210} = \sin~\theta,
\nonumber\\
f^{\prime}_{H11} = f^{\prime}_{H22} = \frac{\gamma}{2},\quad
f^{\prime}_{H33} = \frac{\gamma^{2}}{2cos~\theta}, \quad
\nonumber \\
f^{\prime}_{K_{2}11} = - f^{\prime}_{K_{2}22} = f^{\prime}_{H_{2}12} = \frac{\gamma^{2} - 1}{2m_{1}},
\nonumber \\
f^{\prime}_{H_{n+1}H_{n}1} = f^{\prime}_{K_{n+1}H_{n}1} = f^{\prime}_{H_{n+1}K_{n}2} = - f^{\prime}_{K_{n+1}H_{n}2} 
=\alpha_{n+1}
\nonumber\\
f^{\prime}_{K_{n}H_{n}3} = \frac{\gamma^{2} - 2n\cos\theta}{a\cos\theta}~\alpha_{n}
\eea
The scalar + Yang-Mills part of the Lagrangian 
$
T_{s + YM}
$
has three parts: One is generalizing the preceding
expression (\ref{LagrangeanSM})  with some minor change of the coefficients:
\bea
T_{s+YM}^{(1)} = 
\sin\theta [ ( \Phi_{2}~\phi_{1\mu} - \Phi_{1}~\phi_{2\mu})~v_{0}^{\mu} 
+ m_{1}~( \Phi_{2}~\tilde{u}_{1} - \Phi_{1}~\tilde{u}_{2})~u_{0}
\nonumber\\
+ \frac{\gamma}{2}~[ ( \Phi_{3}~\phi_{2\mu} - \Phi_{2}~\phi_{3\mu})~v_{1}^{\mu} 
+ (m_{2}~\Phi_{3}~\tilde{u}_{2} - m_{3} \Phi_{2}~\tilde{u}_{3})~u_{1}
\nonumber\\
+ ( \Phi_{1}~\phi_{3\mu} - \Phi_{3}~\phi_{1\mu})~v_{2}^{\mu} 
+ (m_{3}~\Phi_{1}~\tilde{u}_{3} - m_{1} \Phi_{3}~\tilde{u}_{1})~u_{2} ]
\nonumber\\
+ \frac{2\cos^{2}\theta - \gamma^{2}}{2\cos\theta}~[ ( \Phi_{2}~\phi_{1\mu} - \Phi_{1}~\phi_{2\mu})~v_{3}^{\mu} 
+ m_{1}~( \Phi_{2}~\tilde{u}_{1} - \Phi_{1}~\tilde{u}_{2})~u_{3} ]
\nonumber\\
+ \frac{\gamma}{2}~[ ( \Phi_{H}~\phi_{1\mu} - \Phi_{1}~\partial_{\mu}\Phi_{H})~v_{1}^{\mu} 
+ m_{1}~\Phi_{H}~\tilde{u}_{1}~u_{1}
\nonumber\\
+ ( \Phi_{H}~\phi_{2\mu} - \Phi_{2}~\partial_{\mu}\Phi_{H})~v_{2}^{\mu} 
+ m_{1}~\Phi_{H}~\tilde{u}_{2}~u_{2} ]
\nonumber\\
+\frac{\gamma^{2}}{2\cos\theta}~[ ( \Phi_{H}~\phi_{3\mu} - \Phi_{3}~\partial_{\mu}\Phi_{H})~v_{3}^{\mu} 
+ m_{3}~\Phi_{H}~\tilde{u}_{3}~u_{3} ].
\eea
The other two parts of
$
T_{s + YM}
$
contain the new scalar fields and will be not given in detail here but can be obtained from the 
expressions
$
f^{\prime}_{abc}
$
listed above.

In the third case, corresponding to
$
\gamma < 1
$
we have 
$
I_{3} = \{ H, K, H_{1}, K_{1} \}
$

The non-zero expressions
$
f^{\prime}_{abc}
$
from the second line of the interaction Lagrangian (\ref{T-sm}) are:
\bea
f^{\prime}_{321} = - f'_{312} = \frac{\gamma}{2}, \quad
f^{\prime}_{123} = - \frac{2\cos^{2}\theta - \gamma^{2}}{2cos~\theta},\quad 
f^{\prime}_{210} = \sin~\theta,
\nonumber\\
f^{\prime}_{H11} = f^{\prime}_{H22} = \frac{3\gamma^{2}\beta m_{1}}{4},\quad
f^{\prime}_{H33} = \frac{\beta m_{3}}{4 \alpha}, \quad
\nonumber \\
f^{\prime}_{K11} = f^{\prime}_{K22} = \frac{1}{m_{1}(3 \gamma^{2} + 1)},\quad
f^{\prime}_{K33} =  \frac{m_{3}}{m_{1}^{2}(3 \gamma^{2} + 1)}
\nonumber\\
f^{\prime}_{K13} = f^{\prime}_{H_{1}23} =  \frac{\gamma\beta m_{1}}{2},  \quad
f^{\prime}_{H_{1}H1} = f^{\prime}_{K_{1}H2} = \alpha, \quad
f^{\prime}_{K_{1}H_{1}3} = \lambda 
\eea
The scalar + Yang-Mills part of the Lagrangian 
$
T_{s + YM}
$
has three parts: One is generalizing the preceding
expression with some minor change of the coefficients:
\bea
T_{s+YM}^{(1)} = 
\sin\theta [ ( \Phi_{2}~\phi_{1\mu} - \Phi_{1}~\phi_{2\mu})~v_{0}^{\mu} 
+ m_{1}~( \Phi_{2}~\tilde{u}_{1} - \Phi_{1}~\tilde{u}_{2})~u_{0}
\nonumber\\
+ \frac{\gamma}{2}~[ ( \Phi_{3}~\phi_{2\mu} - \Phi_{2}~\phi_{3\mu})~v_{1}^{\mu} 
+ (m_{2}~\Phi_{3}~\tilde{u}_{2} - m_{3} \Phi_{2}~\tilde{u}_{3})~u_{1}
\nonumber\\
+ ( \Phi_{1}~\phi_{3\mu} - \Phi_{3}~\phi_{1\mu})~v_{2}^{\mu} 
+ (m_{3}~\Phi_{1}~\tilde{u}_{3} - m_{1} \Phi_{3}~\tilde{u}_{1})~u_{2} ]
\nonumber\\
+ \frac{2\cos^{2}\theta - \gamma^{2}}{2\cos\theta}~[ ( \Phi_{2}~\phi_{1\mu} - \Phi_{1}~\phi_{2\mu})~v_{3}^{\mu} 
+ m_{1}~( \Phi_{2}~\tilde{u}_{1} - \Phi_{1}~\tilde{u}_{2})~u_{3} ]
\nonumber\\
+ \frac{3\gamma^{2}\beta m_{1}}{4\alpha}~[ ( \Phi_{H}~\phi_{1\mu} - \Phi_{1}~\partial_{\mu}\Phi_{H})~v_{1}^{\mu} 
+ m_{1}~\Phi_{H}~\tilde{u}_{1}~u_{1}
\nonumber\\
+ ( \Phi_{H}~\phi_{2\mu} - \Phi_{2}~\partial_{\mu}\Phi_{H})~v_{2}^{\mu} 
+ m_{1}~\Phi_{H}~\tilde{u}_{2}~u_{2} ]
\nonumber\\
+\frac{\beta m_{3}}{4 \alpha}~[ ( \Phi_{H}~\phi_{3\mu} - \Phi_{3}~\partial_{\mu}\Phi_{H})~v_{3}^{\mu} 
+ m_{3}~\Phi_{H}~\tilde{u}_{3}~u_{3} ]
\nonumber \\
+ \frac{1}{m_{1}(3 \gamma^{2} + 1)}~[ ( \Phi_{K}~\phi_{1\mu} - \Phi_{1}~\partial_{\mu}\Phi_{K})~v_{1}^{\mu} 
+ m_{1}~\Phi_{K}~\tilde{u}_{1}~u_{1}
\nonumber\\
+ ( \Phi_{K}~\phi_{2\mu} - \Phi_{2}~\partial_{\mu}\Phi_{K})~v_{2}^{\mu} 
+ m_{2}~\Phi_{K}~\tilde{u}_{2}~u_{2} ]
\nonumber\\
+\frac{m_{3}}{m_{1}^{2}(3 \gamma^{2} + 1)}~[ ( \Phi_{K}~\phi_{3\mu} - \Phi_{3}~\partial_{\mu}\Phi_{K})~v_{3}^{\mu} 
+ m_{3}~\Phi_{K}~\tilde{u}_{3}~u_{3} ].
\eea

The general solution of the gauge invariance problem (in the second order of the perturbation theory)
is given by a direct sum between one of the three solutions described above and an extra piece of the form
\bea
T = (S_{a})_{jk} \Phi_{j}~\partial_{\mu}\Phi_{k}~v_{a}^{\mu}.
\label{S}
\eea
where
$
S_{a}~a = 0,\dots,3
$ 
is a representation of the gauge algebra.
\newpage
\section{Conclusions}

We have seen that gauge invariance in the first two orders of the perturbation theory for Yang-Mills
models leads to a problem of classification of symplectic representations i.e. real, antisymmetric (and
irreducible) representations of the Lie algebra relevant for the model; in our case
$
u(1) \otimes su(2) \simeq u(1) \otimes so(3)
$.

From the physical point of view it follows from above that in the case of a single Higgs field i.e.
$
|I_{3}| = 1
$
we get exactly the Yang-Mills Lagrangian of the standard model. In the general case we have much more
solutions and we need a way to select the physical ones.

%\newpage

\end{document}